%% file: paper.tex
\documentclass[aps,pre,twocolumn,amsmath,amssymb,superscriptaddress,showpacs,longbibliography]{revtex4-1}

\usepackage{graphicx}
\usepackage{dcolumn}
\usepackage{bm}
\usepackage{mathrsfs}

\usepackage{natbib}
\usepackage[text={7in,9.5in},centering]{geometry}

\usepackage{hyperref}
\usepackage{url}

\usepackage{epsfig}
\usepackage{amsmath}
\usepackage{amssymb}
\usepackage{textcomp}

\usepackage{color}
\usepackage{float}

\usepackage{stackengine}
\usepackage{verbatim}

\begin{document}

\title{Simple estimation of hierarchical positions and uncertainty in networks of asymmetric interactions}

\author{G. Tim\'ar}
 \email[]{gtimar@ua.pt}
 \affiliation{Departamento de F\'\i sica da Universidade de Aveiro \& I3N, Campus Universit\'ario de Santiago, 3810-193 Aveiro, Portugal}



\date{\today}

\begin{abstract}
In many social networks it is a useful assumption that, regarding a given quality, an underlying hierarchy of the connected individuals exists, and that the outcome of interactions is to some extent determined by the relative positions in the hierarchy. We consider a simple and broadly applicable method of estimating individual positions in a linear hierarchy, and the corresponding uncertainties. The method relies on solving a linear system characterized by a modified Laplacian matrix of the underlying network of interactions, and is equivalent to finding the equilibrium configuration of a system of directed linear springs. We provide a simple first-order approximation to the exact solution, which can be evaluated in linear time. The uncertainty of the hierarchy estimate is determined by the network structure and the potantial energy of the corresponding spring system in equilibrium. The method generalizes straightforwardly to multidimensional hierarchies and higher-order, non-pairwise interactions.
\end{abstract}

\maketitle

\section{Introduction}
\label{sec1}

Interactions between individuals in networked systems are often asymmetric, necessitating a directed network description. In many situations it is useful to assume that there is an underlying hierarchy of the interacting nodes, which may be inferred from the structure of the network. Such hierarchy may refer to the trophic levels in food webs \cite{williams2000simple}, dominance hierarchies in animal groups \cite{de1998finding}, or organizational hierarchies in social and technological networks \cite{pumain2006hierarchy}. There is a wide range of contexts in which hierarchical structures appear, and consequently a whole host of methods exists to infer hierarchy in each particular type of system \cite{trusina2004hierarchy, clauset2008hierarchical, maiya2009inferring, gupte2011finding, shizuka2012social, corominas2013origins, zhang2013potential, clauset2015systematic, czegel2015random}.

With regards to dynamical processes on directed networks a useful hierarchical description relies on the concept of reachability between nodes, and nodes are characterized mainly by their relative positions with respect to sources, sinks and strongly connected components \cite{mones2012hierarchy, timar2017mapping, wright2019central}. In this case standard eigenvector-based centrality measures, such as the original eigenvector centrality \cite{bonacich1987power}, Katz centrality \cite{bonacich2001eigenvector}, or PageRank \cite{brin1998anatomy}, may also be appropriate. In a more structural context, often it may be assumed that nodes belong to hierarchical levels and directed links predominantly point from lower to higher level nodes, e.g., in Twitter follow networks \cite{myers2014information}. Working with this assumption, a penalty function may be introduced for links pointing in the wrong direction. A number of studies have defined hierarchy as the assignment of levels in the network that minimizes the sum total of such penalties, and have provided efficient algorithms to find this assignment \cite{gupte2011finding, tatti2014faster, tatti2015hierarchies, letizia2018resolution}. These methods are closely related with the minimum feedback arc set problem, which has also been used directly to infer hierarchies in networks \cite{zhao2017feedback}. In these works directed links from lower to higher levels may connect nodes of arbitrary level difference, meaning that a directed acyclic graph necessarily has a penalty-free level assignment. Dealing with level inconsistencies in directed acyclic graphs may also be important in certain contexts, e.g., in most food webs trophic levels cannot be unambiguously assigned such that they are all integer values. (Such an assignment can only be made if any traversal of any cycle in the network finds an equal number of directed links with and against the direction of traversal.) Commonly used methods of resolving these inconsistencies in the context of food webs can be found in Ref. \cite{williams2004limits}, but there is no generally accepted approach.

Often link directions themselves are unknown and have to be inferred from structural properties of the equivalent undirected network. A natural approach to do this is to compare the degrees of neighbouring nodes and assign link directions such that links always point in the direction of the higher degree node \cite{trusina2004hierarchy}. Another method treats the observed undirected network as a result of a generative process given a hierarchy (a rooted spanning tree) and provides a greedy algorithm to generate the maximum likelihood hierarchy \cite{maiya2009inferring}.

More frequently link directions are inferred from the results of certain quantifiable interactions between pairs of connected nodes. These may be the binary outcome of dominance interactions (winner/loser) in a group of individuals, results of competitive matchups or any measurable result of relative performance.
Literature on establishing a ranking of items/individuals based on pairwise comparisons dates back to Thurstone who focused on comparisons of the perception of psychophysical stimuli and social values \cite{thurstone1927law, thurstone1927psychophysical, thurstone1927method}. He assumed that multiple evaluations of the same stimulus may be placed on a continuous scale such that they follow a normal distribution with a well-defined mean and standard deviation. This model has been generalized in various directions. Some notable examples include, e.g., comparisons of multiple (more than two) items \cite{vojnovic2016parameter}, varying covariance terms for discriminal processes \cite{takane1981maximum} and mutliple choices \cite{orban2019generalization}.
A second main branch of research on pairwise comparisons is based on the Bradley-Terry framework \cite{bradley1952rank}, where it is assumed that each item/individual $i$ has a unique positive number $\gamma_i$ assigned to it, and the probability that $i$ ``beats'' $j$ in a comparison is $\gamma_i / (\gamma_i + \gamma_j)$. This framework has been used in conjunction with a binomial model of multiple interactions to estimate animal dominance hierarchies \cite{boyd1983method, mcmahan1984application}. An iterative method for maximum likelihood estimation in a wide class of generalization of the Bradley-Terry model can be found in \cite{hunter2004mm}. A more recent, intuitive, random-walk-based method is given in \cite{negahban2012iterative}.
Methods for estimation from pairwise comparisons are widely used in designing recommender systems \cite{park2012literature, portugal2018use} or skill rating systems such as the Elo system in chess \cite{elo1978rating} or the more recent TrueSkill \cite{herbrich2006trueskill}.

In this paper we consider a simple and broadly applicable approach to estimate a linear hierarchy from an arbitrary connected network of quantifiable interactions. We assume that the interaction results indicate a difference in performance of the two interacting individuals. Following other Thurstone-type models we assume that the performance of an individual is a normally distributed random variable. We show that the maximum likelihood hierarchy is related with a generalized Laplacian matrix of the network of interactions. The problem is equivalent to finding the equilibrium state of a system of directed linear springs. This setting makes it possible to obtain the uncertainty of individual hierarchical position estimates in a straightforward, natural manner---contrary to most of the existing methods for hierarchy estimation. In our method the uncertainty in the hierarchy estimation is directly related with the potantial energy of the corresponding linear spring system in equilibrium.

The rest of the paper is structured as follows. In Sec. \ref{sec2} we introduce the normal generative model for interaction outcomes and show how this may be used as an approximation to a sometimes more appropriate binomial model of outcomes. In Sec. \ref{sec3} we derive the maximum likelihood estimate of hierarchical positions using a generalized Laplacian matrix of the underlying network. In Sec. \ref{sec4} we calculate the individual uncertainties of the hierarchical position estimates. As an example, in Sec. \ref{sec5} we apply our method to the network of 21st-century professional male tennis players. We give our conclusions in Sec. \ref{sec6}.

\section{Normal model of interaction outcomes}
\label{sec2}

We consider a system of $N$ nodes, where some pairs of nodes engage in interactions. This information is represented in the adjacency matrix: $A_{ij} = A_{ji}$ is the number of interactions between nodes $i$ and $j$. Interactions must occur between two different nodes, i.e., no self-loops are allowed: $A_{ii} = 0$ for all $i$. The total number of interactions is equal to the number of links in the network, $L = \sum_{i<j} A_{ij}$. We assume that each node $i$ has a hierarchical position $h_i$ associated with it, and that node $i$'s ``performance'' in a given interaction is a normally distributed random variable with mean $h_i$ and variance $c v_i$. (The $v_i$ are assumed to be known node-specific values, while the constant factor $c$ is to be estimated.) The choice of a normal model for performance is widely used, e.g., in the popular Elo and TrueSkill ranking systems. If we assume that performance is a result of the aggregation of many independent random effects, then a normal distribution naturally arises as a consequence of the central limit theorem. Additionally, the normal distribution is the maximum entropy, i.e., least informative distribution with a given mean and variance, which provides further justification for its use. The normal model also arises as an approximation to certain other models of interactions, e.g. a binomial model of multiple interactions (see Sec. \ref{sec21}).

Let $I_i$ denote the $i$-th interaction, and let $I_i(1), I_i(2)$ denote the indices of the two nodes participating in the interaction. We assume that the result of the $i$-th interaction is a random variable $r_i = \rho_{I_i(1)} - \rho_{I_i(2)}$, where $\rho_k$ denotes the instantaneous performance of node $k$. The result $r_i$ is therefore also a normally distributed random variable,

\begin{align}
P(r_i|h_{I_i(1)}, h_{I_i(2)}, c) = \frac{1}{\sqrt{2 \pi c V_i}} e^{ -\frac{1}{2}   \frac{ \left( h_{I_i(1)} - h_{I_i(2)} - r_i  \right)^2 }{c V_i }    },
\label{eq200.10}
\end{align}

\noindent
where $V_i = v_{I_i(1)} + v_{I_i(2)}$. Often it might be more appropriate to treat the $V_i$ directly as the known parameters and not the $v_i$, as we may have more information about the uncertainty of interaction outcomes than of individual performances. Note that interactions $I_i$ and $I_j$ are equivalent if $I_i(1) = I_j(2)$, $I_j(1) = I_i(2)$, and $r_i = -r_j$.
Given the set of $r_i$ and $V_i$ values we wish to infer the hierarchical positions $h_i$ and the constant $c$. The joint distribution for the complete vector $\mathbf{r}$ of observed results given the vector $\mathbf{h}$ of positions and $c$ is written as

\begin{align}
P(\mathbf{r}|\mathbf{h}, c) = \prod_{i=1}^L \left[ \frac{1}{ \sqrt{2 \pi c V_i}} e^{ -\frac{1}{2}  \frac{ \left( h_{I_i(1)} - h_{I_i(2)} - r_i \right)^2 }{c V_i}   }   \right].
\label{eq200.20}
\end{align}

\subsection{Normal approximation to a binomial model of interaction outcomes}
\label{sec21}

The normal model of Sec. \ref{sec2} may also be derived as an approximation to certain other models of interaction outcomes. Let us consider a discrete generative model, where pairs of nodes interact multiple times and each interaction results in a clear winner. We assume that for each encounter only the binary result of who won/lost is known and not by what margin. We also assume that in each encounter the difference in the probabilities of winning is given by the difference in hierarchical position, i.e., $p_i - p_j = h_i - h_j$. This leads to

\begin{align}
p_i = \frac{h_i - h_j + 1}{2}, \quad \quad p_j = \frac{h_j - h_i + 1}{2}.
\label{eq210.10}
\end{align}

\noindent
The number of encounters between nodes $i$ and $j$ is represented by $A_{ij}$ as before.
The result at the end of the $A_{ij}$ encounters for a given pair is an integer number $R_{ij} = w_i - w_j$, the difference in the number of wins.
Assuming that all encounters are independent, the result $R_{ij}$ is a discrete random variable with a binomial distribution,

\begin{align}
p(R_{ij}) = \binom{A_{ij}}{ \frac{R_{ij} + A_{ij}}{2} } p_i^{ \frac{R_{ij} + A_{ij}}{2} } p_j^{A_{ij} - \frac{R_{ij} + A_{ij}}{2}}.
\label{eq210.20}
\end{align}

\noindent
Assuming that the number of wins of either node is large enough, this may be approximated with a normal distribution,

\begin{align}
p(R_{ij}) \approx \frac{1}{\sqrt{ 2 \pi A_{ij} p_i p_j }} e^{ -\frac{1}{2}  \frac{ \left( \frac{R_{ij} + A_{ij}}{2} - A_{ij} p_i \right)^2 }{ A_{ij} p_i p_j } }.
\label{eq210.30}
\end{align}

\noindent
Let us further assume that $p_i \approx p_j$, i.e., that $|p_i - 1/2| = |p_j - 1/2| \ll 1$. Then we can write

\begin{align}
p(R_{ij}) \approx  \sqrt{ \frac{2}{\pi A_{ij}} } \, e^{ -2  \frac{ \left( \frac{R_{ij} + A_{ij}}{2} - A_{ij} p_i \right)^2 }{ A_{ij} } }.
\label{eq210.40}
\end{align}

\noindent
Using Eq. (\ref{eq210.10}) we get

\begin{align}
p(R_{ij}) \approx  \sqrt{ \frac{2}{\pi A_{ij}} } \, e^{ -  \frac{A_{ij}}{2}  \left( h_i - h_j -   \frac{R_{ij}}{A_{ij}}   \right)^2   }.
\label{eq210.50}
\end{align}

\section{Maximum likelihood estimate}
\label{sec3}

The function in Eq. (\ref{eq200.20}) may be treated as a likelihood function for the parameters $\mathbf{h}$: $\mathcal{L}(\mathbf{h}) \equiv P(\mathbf{r}|\mathbf{h}, c)$. It is immediately clear that this function has the translational symmetry $\mathcal{L}(\mathbf{h}) = \mathcal{L}(\mathbf{h} + f \mathbf{1})$, where $f$ is an arbitrary constant and $\mathbf{1}$ is the vector of all ones of dimension $N$. This means that $\mathcal{L}$ does not have a unique maximum, and the hierarchical positions $h_i$ may only be determined up to an additive constant. Let us fix one of the variables, say, $h_u = a$, where $a$ is a constant. This means restricting the likelihood function to the $N-1$ dimensional hyperplane defined by $h_u = a$. The restricted function is a function of only $N-1$ independent variables. Let $\Phi$ denote the set of indices of interactions where neither interacting node is node $u$, and $\Psi$ the set where one of the nodes is node $u$. The restricted likelihood function is then written as

\begin{align}
\mathcal{L}^{(u)}&(\mathbf{h}^{(u)}) = \prod_{i \in \Phi} \left[ \frac{1}{ \sqrt{2 \pi c V_i}} e^{ -\frac{1}{2}  \frac{ \left( h_{I_i(1)} - h_{I_i(2)} - r_i \right)^2 }{c V_i}   }   \right] \times \nonumber \\
&\times \prod_{i \in \Psi} \left[ \frac{1}{ \sqrt{2 \pi c V_i}} e^{ -\frac{1}{2}  \frac{ \left( h_{I_i(1)} - a - r_i \right)^2 }{c V_i}   }   \right],
\label{eq300.10}
\end{align}

\noindent
where we assumed that the node pairs in $\Psi$ are ordered such that the second ineracting node is always node $u$, i.e., $I_i(2) = u$ for all $i \in \Psi$. The vector $\mathbf{h}^{(u)}$ is a reduced $N-1$ dimensional vector whose components are $\{ h_1, \ldots, h_{u-1}, h_{u+1}, \ldots, h_N \}$. Eq. (\ref{eq300.10}) may be written more compactly as

\begin{align}
\mathcal{L}^{(u)}(\mathbf{h}^{(u)}) = \left[ \prod_{i=1}^L  \frac{1}{ \sqrt{2 \pi c V_i}}  \right]  e^{ -\frac{1}{2c}  Q( \mathbf{h}^{(u)} )   },
\label{eq300.20}
\end{align}

\noindent
where $Q( \mathbf{h}^{(u)} )$ is a quadratic function of the form

\begin{align}
Q( \mathbf{h}^{(u)} ) = (\mathbf{h}^{(u)})^T \mathbf{M} \mathbf{h}^{(u)} + \mathbf{b}^T \mathbf{h}^{(u)} + d.
\label{eq300.30}
\end{align}

\noindent
Comparing the corresponding coefficients in Eqs. (\ref{eq300.10}) and (\ref{eq300.30}) we can write the elements of the matrix $\mathbf{M}$ and the vector $\mathbf{b}$,

{
\medmuskip=0mu
\thinmuskip=0mu
\thickmuskip=0mu
\begin{align}
M_{ij} &= -\frac{A_{ij}}{v_{ij}} + \delta_{ij} \sum_k \frac{A_{ik}}{v_{ik}} \label{eq300.35} \\
b_i &= -2 \left[ \sum_k \frac{A_{ik} r_{ik}}{v_{ik}} + a \frac{A_{iu}}{v_{iu}} \right].
\label{eq300.40}
\end{align}
}

\noindent
where $r_{ij}$ denotes the average result of interactions between nodes $i$ and $j$. We also assumed that the variance for all interactions between $i$ and $j$ are equal and denoted by $v_{ij}$. (The solution without this assumption is also straightforward, only the expressions for $M_{ij}$ and $b_i$ would be more complicated.) Note that the indices $i$ and $j$, in Eqs. (\ref{eq300.35}) and (\ref{eq300.40}), may take any integer value between $1$ and $N$, except $u$. The summation, in both Eqs. (\ref{eq300.35}) and (\ref{eq300.40}), however, is over all indices, including $u$.
The matrix $\mathbf{M}$---a weighted, reduced Laplacian \cite{newman2018networks}---is positive definite for arbitrary $v_{ij}$ and arbitrary choice of $u$ if the underlying network (encoded by the adjacency matrix $\mathbf{A}$) is connected (see Appendix for a proof).
The quadratic function $Q( \mathbf{h}^{(u)} )$ may be interpreted as the total potential energy of a system of directed linear springs along a line where spring $i$ has a directed equilibrium length of $r_i$, a stiffness of $2/V_i$, and is imagined to connect the points $h_{I_i(1)}$ and $h_{I_i(2)}$. The maximum likelihood estimate for $\mathbf{h}^{(u)}$ therefore coincides with the minimum energy (equilibrium) configuration of the corresponding spring system. Since $\mathbf{M}$ is positive definite, the position of the minimum of the quadratic function $Q(\mathbf{h}^{(u)})$ is given as

\begin{align}
\mathbf{h}^{(u)*} = -\frac{1}{2} \mathbf{M}^{-1} \mathbf{b}.
\label{eq300.50}
\end{align}

\noindent
Instead of performing the matrix inversion, it is generally more efficient to solve the equivalent linear system

\begin{align}
\mathbf{M} \mathbf{h}^{(u)*} = -\frac{1}{2} \mathbf{b}.
\label{eq300.55}
\end{align}

\noindent
If the underlying network is sparse, then so is matrix $\mathbf{M}$ and the solution of Eq. (\ref{eq300.55}) may be found efficiently using dedicated solvers.

The maximum likelihood estimate for the variable $c$ may be obtained in the standard way, differentiating the log likelihood [using Eq. (\ref{eq300.20})] with respect to $c$ and making it equal to zero. The result is

\begin{align}
c^* = \frac{Q(\mathbf{h}^{(u)*})}{L} = \frac{E_{\textrm{tot}}}{L},
\label{eq300.60}
\end{align}

\noindent
which is the mean potential energy per link of the corresponding spring system at equilibrium.

\subsection{First-order approximation}
\label{sec31}


The linear system (\ref{eq300.55}) may also be solved iteratively, via e.g., the Jacobi method, whereby Eq. (\ref{eq300.55}) is rewritten in the form

\begin{align}
\mathbf{h}^{(u)*}(k+1) = \mathbf{T} \mathbf{h}^{(u)*}(k) + \mathbf{c},
\label{eq310.10}
\end{align}

\noindent
and iterated until convergence. The matrix $\mathbf{T}$ and vector $\mathbf{c}$ are given as

{
\medmuskip=0mu
\thinmuskip=0mu
\thickmuskip=0mu
\begin{align}
T_{ij} &= \frac{A_{ij}}{v_{ij}} \bigg/ \sum_k \frac{A_{ik}}{v_{ik}} \label{eq310.20} \\
c_i &= \left( \sum_j \frac{A_{ij} r_{ij}}{v_{ij}} + a \frac{A_{iu}}{v_{iu}} \right) \bigg/ \sum_k \frac{A_{ik}}{v_{ik}}.
\label{eq310.30}
\end{align}
}

\noindent
The matrix $\mathbf{T}$ is a reduced, normalized, weighted adjacency matrix. All of its rows sum to $1$, except for the rows that correspond to neighbours of node $u$ in the underlying network. The matrix is, therefore, ``almost'' a stochastic matrix, with a largest eigenvalue usually close to $1$, which results in slow convergence in general. We may nonetheless use Eq. (\ref{eq310.10}) to write a first-order approximation to the exact solution, which may be evaluated in linear time. Assuming $\mathbf{h}^{(u)*}(0) = \mathbf{0}$ and $a = 0$ we have, as an approximation to the $i$-th component,

\begin{align}
h^{(u)*}_i \approx \frac{ \sum_j r_{ij} \frac{A_{ij}}{v_{ij}} }{ \sum_j \frac{A_{ij}}{v_{ij}} },
\label{eq310.40}
\end{align}

\noindent
which is a weighted average of the results of node $i$'s interactions. For a simple, unweighted directed network Eq. (\ref{eq310.40}) simplifies to

\begin{align}
h^{(u)*}_i \approx \frac{ q_{\textrm{in}} - q_{\textrm{out}} }{ q_{\textrm{in}} + q_{\textrm{out}} },
\label{eq310.50}
\end{align}

\noindent
i.e., the difference of in- and out-degree divided by the total degree of node $i$.

For relatively dense networks Eq. (\ref{eq310.40}) [or Eq. (\ref{eq310.50})] may provide a reasonable approximation to the exact values, as we demonstrate through an example in Sec. \ref{sec5}.

\section{Uncertainty of the estimate}
\label{sec4}

To obtain the uncertainty related with the estimated variables $h_i$ we begin with Bayes' theorem,

\begin{align}
P ( \mathbf{h}^{(u)}|\mathbf{r} ) = \frac{ P ( \mathbf{r}|\mathbf{h}^{(u)} ) P(\mathbf{h}^{(u)}) }{ P ( \mathbf{r} ) },
\label{eq400.10}
\end{align}

\noindent
where $P ( \mathbf{r}|\mathbf{h}^{(u)} ) = \mathcal{L}^{(u)}(\mathbf{h}^{(u)})$ is the restricted likelihood function. (Note that the posterior distribution $P ( \mathbf{h}^{(u)}|\mathbf{r} )$ may only be defined over the reduced vector $\mathbf{h}^{(u)}$, as the original likelihood function $P ( \mathbf{r}|\mathbf{h} )$ is not integrable due to its translational symmetry.) We only consider the dependence on $\mathbf{h}^{(u)}$, while $c = c^*$ [see Eq. (\ref{eq300.60})] is considered a constant. Assuming a uniform prior distribution $P(\mathbf{h}^{(u)})$, we have that the posterior distribution is proportional to the likelihood,

\begin{align}
P ( \mathbf{h}^{(u)}|\mathbf{r} ) \propto P ( \mathbf{r}|\mathbf{h}^{(u)} ).
\label{eq400.20}
\end{align}

\noindent
Using Eq. (\ref{eq300.20}) in Eq. (\ref{eq300.30}) and considering that the matrix $\mathbf{M}$ is positive definite, we see that the posterior distribution is multivariate normal centered at $\mathbf{h}^{(u)*}$, with covariance matrix $c^* \mathbf{M}^{-1}$:

\begin{align}
P ( \mathbf{h}^{(u)}|\mathbf{r} ) \propto e^{ -\frac{1}{2c^*} ( \mathbf{h}^{(u)} - \mathbf{h}^{(u)*} )^T \mathbf{M} ( \mathbf{h}^{(u)} - \mathbf{h}^{(u)*} )}.
\label{eq400.30}
\end{align}

\noindent
The uncertainty related with $h_i$ can be identified as the standard deviation of the corresponding marginal distribution,

\begin{align}
P_i(h_i) = \int_{-\infty}^{\infty} \ldots & \int_{-\infty}^{\infty} P ( \mathbf{h}^{(u)}|\mathbf{r} ) dh_1  \ldots dh_{i-1} \times \nonumber \\
&\times dh_{i+1} \ldots dh_{N-1}.
\label{eq400.40}
\end{align}

\noindent
The distribution $P_i(h_i)$, however, depends on the choice of $h_u$ (the variable to keep fixed) and is therefore not appropriate to characterize the uncertainty of individual estimates. To resolve this issue, instead of fixing $h_u = a$, let us restrict the original likelihood function [Eq. (\ref{eq200.20})] to the $N-1$ dimensional hyperplane defined by the condition $\sum_i h_i = \tilde{a}$, where $\tilde{a}$ is a constant. This family of hyperplanes is the only symmetric choice, as these are the only hyperplanes perpendicular to the direction of translational symmetry. The restricted function is a function of $N-1$ independent variables, and $h_u$ is now expressed with the others as $h_u = \tilde{a} - \sum_{i \neq u} h_i$. Similarly to Eq. (\ref{eq300.10}), the restricted likelihood may be written as

\begin{align}
\mathcal{L}^{(u)}&(\mathbf{h}^{(u)}) = \prod_{i \in \Phi} \left[ \frac{1}{ \sqrt{2 \pi c V_i}} e^{ -\frac{1}{2}  \frac{ \left( h_{I_i(1)} - h_{I_i(2)} - r_i \right)^2 }{c V_i}   }   \right] \times \nonumber \\
&\times \prod_{i \in \Psi} \left[ \frac{1}{ \sqrt{2 \pi c V_i}} e^{ -\frac{1}{2}  \frac{ \left( h_{I_i(1)} - (\tilde{a} - \sum_{k(\neq u)} h_k) - r_i \right)^2 }{c V_i}   }   \right],
\label{eq400.50}
\end{align}

\noindent
or more compactly as

\begin{align}
\mathcal{L}^{(u)}(\mathbf{h}^{(u)}) = \left[ \prod_{i=1}^L  \frac{1}{ \sqrt{2 \pi c V_i}}  \right]  e^{ -\frac{1}{2c}  \tilde{Q}( \mathbf{h}^{(u)} )   },
\label{eq400.60}
\end{align}

\noindent
where $\tilde{Q}( \mathbf{h}^{(u)} )$ is a quadratic function of the form

\begin{align}
\tilde{Q}( \mathbf{h}^{(u)} ) = (\mathbf{h}^{(u)})^T \tilde{\mathbf{M}} \mathbf{h}^{(u)} + \tilde{\mathbf{b}}^T \mathbf{h}^{(u)} + d.
\label{eq400.70}
\end{align}

\noindent
Comparing the corresponding coefficients in Eqs. (\ref{eq400.50}) and (\ref{eq400.70}) we can write the elements of the matrix $\tilde{\mathbf{M}}$ and the vector $\tilde{\mathbf{b}}$,

{
\medmuskip=0mu
\thinmuskip=0mu
\thickmuskip=0mu
\begin{align}
\tilde{M}_{ij} &= -\frac{A_{ij}}{v_{ij}} + \delta_{ij} \sum_k \frac{A_{ik}}{v_{ik}} + \frac{A_{iu}}{v_{iu}} + \frac{A_{ju}}{v_{ju}} + \sum_k \frac{A_{ku}}{v_{ku}} \label{eq400.35} \\
\tilde{b}_i &= -2 \left[ \sum_j \frac{A_{ij} r_{ij}}{v_{ij}} + \sum_j \frac{A_{ju} (\tilde{a} + r_{ju})}{v_{ju}} + \tilde{a} \frac{A_{iu}}{v_{iu}} \right].
\label{eq400.80}
\end{align}
}

\noindent
As before, $r_{ij}$ denotes the average result of interactions between nodes $i$ and $j$ and the variance for all interactions between $i$ and $j$ are considered equal and denoted by $v_{ij}$. 
The solutions for $\mathbf{h}^{(u)}$ and $c$ are given as

\begin{align}
\mathbf{h}^{(u)*} = -\frac{1}{2} \tilde{\mathbf{M}}^{-1} \tilde{\mathbf{b}}, \label{eq400.45} \\
c^* = \frac{\tilde{Q}(\mathbf{h}^{(u)*})}{L} = \frac{E_{\textrm{tot}}}{L}.
\label{eq400.90}
\end{align}

\noindent
The remaining component, $h_u$, can be obtained from the hyperplane condition: $h_u^* = \tilde{a} - \sum_k h_k^{(u)*}$. We note that the solutions $\mathbf{h}^{(u)*}$ are identical, up to an additive constant, to the ones found by using $\mathbf{M}$ and $\mathbf{b}$ [Eq. (\ref{eq300.50})] instead of $\tilde{\mathbf{M}}$ and $\tilde{\mathbf{b}}$. There is a one-to-one correspondence between the values $a$ and $\tilde{a}$ in the sense that the solution found via Eq. (\ref{eq300.50}) using $a$ is identical to the solution found via Eq. (\ref{eq400.45}) using

\begin{align}
\tilde{a} = a - \frac{1}{2} \mathbf{1}^T \mathbf{M}^{-1} \mathbf{b}.
\label{eq400.95}
\end{align}

The resulting posterior distribution over the $h_i$ again has the form of Eq. (\ref{eq400.30}), with matrix $\mathbf{M}$ replaced with $\tilde{\mathbf{M}}$, i.e., multivariate normal with covariance matrix $c^* \tilde{\mathbf{M}}^{-1}$. Due to symmetry, the marginal distributions are now independent of the choice of $h_u$. In particular, $P_i(h_i)$ is a normal distribution with variance $c^*(\tilde{\mathbf{M}}^{-1})_{ii}$. The uncertainty related with the variable $h_i$, i.e., the standard deviation of the corresponding marginal distribution, is given by

\begin{align}
s_i = \sqrt{ \frac{E_{\textrm{tot}}}{L} (\tilde{\mathbf{M}}^{-1})_{ii} },
\label{eq400.100}
\end{align}

\noindent
for all $i \neq u$. To obtain the uncertainty related with $h_u$, we must choose another variable, say, $h_m$, to be the dependent variable and use Eq. (\ref{eq400.100}) to get $s_u$. The matrix $\tilde{\mathbf{M}}$ thus enables us to calculate the uncertainties related with the individual estimates, in addition to the maximum likelihood estimates themselves. The downside is that $\tilde{\mathbf{M}}$ is a dense matrix and the matrix inversion in Eq. (\ref{eq400.100}), although still polynomial time, may become prohibitively slow in large networks.

Note that if the underlying network of interactions is a tree, then there can be no inconsistencies in the results, i.e., $E_{\textrm{tot}}=0$ and all uncertainties are zero: $s_i=0$. This is true even for non-trees, if the results are balanced such that for any cycle, the sum of results (with the correct signs, according to the direction of traversal) is zero. If any cycle in the network is ``imbalanced'', however, this results in a nonzero uncertainty for the position estimates of all nodes in the network.

\section{Example: Estimating tennis players' relative abilities}
\label{sec5}

\begin{figure}[H]
\centering
\includegraphics[width=\columnwidth,angle=0.]{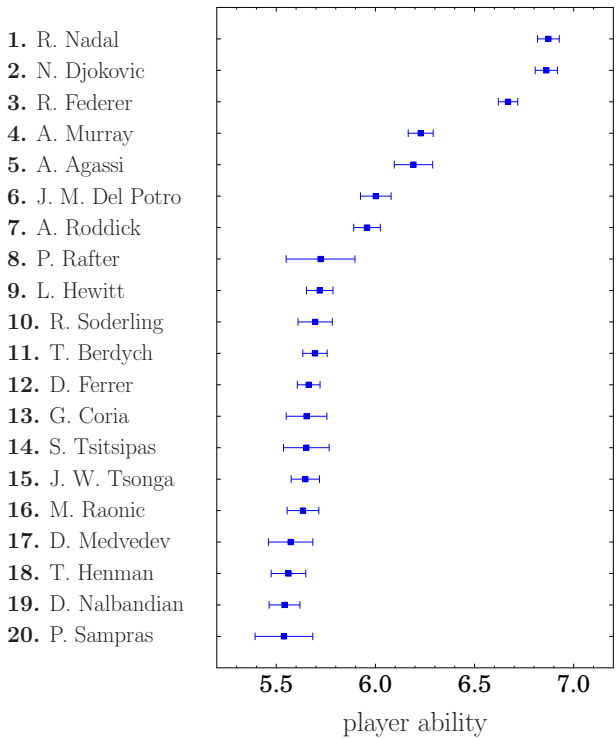}
\caption{Estimate of the abilities of the top 20 professional male tennis players based on all ATP tennis matches played from January 2000 to July 2021. The estimated positions (solid blue squares), calculated using Eq. (\ref{eq400.45}), are relative to the lowest positioned player in this period. The uncertainties [Eq. (\ref{eq400.100})] are represented by horizontal error bars.}
\label{fig:tennis_normal}
\end{figure}

As an illustrative example we consider the network of modern-day professional male tennis players based on all ATP (Association of Tennis Professionals) tennis matches played from January 2000 to July 2021. The data was downloaded from \cite{tennisdata}. We opted for the lowest possible level of analysis given the available data: we considered each finished set, played between players $i$ and $j$, to be one interaction. The difference in the number of games won, for the given set, was considered to be the result of the interaction. All interaction outcomes, for all pairs of players, were assumed to have the same inherent uncertainty, i.e., $v_{ij} = 1$ for all $i,j$ in the expressions for $\mathbf{M}, \mathbf{b}$ [Eqs. (\ref{eq300.35}, \ref{eq300.40})] or $\tilde{\mathbf{M}}, \tilde{\mathbf{b}}$ [Eqs. (\ref{eq400.35}, \ref{eq400.80})]. In these equations the quantity $r_{ij}$ corresponds to the average interaction result over all $A_{ij}$ interactions between players $i$ and $j$. Players that have played less than 10 sets in the given time interval were removed to avoid unrealistic ranking positions due to fluke results. We were left with $N = 896$ players that have played $L = 142570$ sets between them. Using the methods in Sections \ref{sec3} and \ref{sec4} we calculated the estimated position of each player and the uncertainty related with the estimate. According to our definition of interaction result, the difference in hierarchical position between two players approximately represents the expected difference in the number of games won by each if they were to play a set against each other. Fig. \ref{fig:tennis_normal} shows the estimated positions and the uncertainties for the top 20 players in the considered period.

\noindent
All the sets played between players were considered to be of equal importance in the estimate, so these results may represent the approximate relative status of players in terms of pure technical ability. By treating all encounters as equally important, we disregard the mental aspect of the game, e.g., being able to perform well in important matches such as tournament finals, which contribute more to actual ATP rankings and the popular perception of the quality of a player.

\begin{figure}[H]
\centering
\includegraphics[width=\columnwidth,angle=0.]{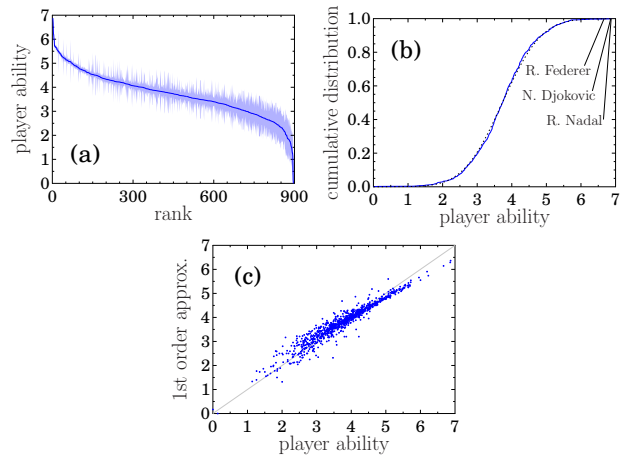}
\caption{(a) Tennis players' ability as a function of rank, considering all $896$ players of the network. Shaded light blue region indicates the uncertainty of the estimate for a given rank. (b) Cumulative distribution of player ability, fitted with a normal distribution (dashed black line). The positions of the top three players are explicitly shown. (c) First-order approximation [calculated using Eq. (\ref{eq310.40})] as a function of the exact estimates of players' abilities.}
\label{fig:tennis_3}
\end{figure}

The complete distribution of hierarchical positions is presented in Fig. \ref{fig:tennis_3}. Fig. \ref{fig:tennis_3}(a) shows the estimated positions as a function of ranking, and the related uncertainty as a shaded light blue region. A clear trend can be observed: lower ranked players have a greater uncertainty related with their estimated positions. This phenomenon (already noticable in the case of the top 20 players, Fig. \ref{fig:tennis_normal}) is due to the greatly varying number, and varying profile of matches played. First, higher quality players have a considerably higher number of interactions, allowing for more accuracy in the estimation of their ability. Secondly, the interactions of high quality players disproportionately involve other high quality players (due to the single elimination structure of tennis tournaments) whose positions are also accurately estimated.

\noindent
Fig. \ref{fig:tennis_3}(b) demonstrates that the distribution of estimated positions can be accurately fitted with a normal distribution using the mean $\mu \approx 3.72$ and standard deviation $\sigma \approx 0.89$ of the data. The highest ranked player, Rafael Nadal, has an estimated position of $\approx 6.87$, which is approximately $3.5$ standard deviations above the average. In Fig. \ref{fig:tennis_3}(c) we show the first-order approximation of Eq. (\ref{eq310.40}) as a function of the exact position estimates. The approximation is reasonably accurate, with a Pearson correlation coefficient of $\sigma \approx 0.96$.

\section{Discussion and conclusions}
\label{sec6}

In this paper we have proposed a simple, transparent and broadly applicable method of estimating the positions of individuals along a linear hierarchy, given a connected network of interaction outcomes. The method relies on the commonly used assumption that the performance---in an abstract sense, potentially---of an individual in any given interaction, is a random variable that is normally distributed around a well-defined mean: the hierarchical position to be estimated. The interaction outcomes are considered to be the difference in the instantaneous performances of the interacting individuals, and are therefore also normally distributed random variables. We have discussed the framework for calculating the maximum likelihood estimate for the hierarchical positions, and have shown that this is always possible if the underlying network of interactions is connected. The problem is equivalent to finding the equilibrium configuration of a network of directed linear springs along a line. Therefore, it reduces to solving a system of linear equations determined by a matrix analogous to the graph Laplacian.

The linear approach we have outlined deals simply and naturally with the issue of inconsistencies in the network of interaction results. It also enables us to calculate, in a straightforward manner, the uncertainty of the estimates for individual hierarchical positions. A drawback of the linear approach is time complexity. Estimating the positions is equivalent to solving a sparse linear system, given that the underlying network is sparse. Using dedicated sparse solvers can significantly reduce the time complexity from the strict upper bound of $\sim O(N^3)$, however, is still superlinear in general. Calculating the uncertainties related with the estimates required the inversion of a dense matrix, leaving no simple option to overcome the limitations caused by time complexity issues. The methods we have presented work well for networks of sizes up to $N \sim 10^4$, but may be prohibitively slow for larger networks. We provided a first-order approximation to the exact position estimates, that can be evaluated in linear time, and hence allow for the analysis of large networks.

Apart from simplicity and transparency, the linear approach has another advantage which is its great flexibility. Multiple types of interactions, or home advantage are easily incorporated into this framework. It is also straightforward to generalize to multidimensional hierarchy, where interactions may also compare different aspects of the qualities of individuals. The maximum likelihood estimates would still depend on the solution of linear systems, albeit with more complicated matrices. One might also imagine higher-order, non-pairwise interactions, or indeed a generalized linear interaction, where the result of the interaction is an arbitrary linear combination of the performances of the participants. This could include situations such as cooperation within groups of individuals and/or competition between groups.

\section*{Acknowledgments}

This work was developed within the scope of the project i3N, UIDB/50025/2020 \& UIDP/50025/2020, financed by national funds through the FCT/MEC--Portuguese Foundation for Science and Technology. G. T. was supported by FCT Grant No. CEECIND/03838/2017.

\subsection*{Appendix: Proof that matrix $\mathbf{M}$ is positive definite}
\label{secB}
\setcounter{equation}{0}
\renewcommand{\theequation}{A\arabic{equation}}

Here we show a proof that the symmetric matrix $\mathbf{M}$ [Eq. (\ref{eq300.35})] is positive definite for arbitrary positive $v_{ij}$ and an arbitrary choice of $u$, if the underlying network is connected. Matrix $\mathbf{M}$ is a weighted, reduced Laplacian matrix, that can be written in the form

\begin{align}
\mathbf{M} = \mathbf{B} \mathbf{B}^T,
\label{eqA100.10}
\end{align}

\noindent
where $\mathbf{B}$ is a weighted, reduced edge incidence matrix of size $N-1 \times L$, whose elements are $B_{ie} = A_{ij} / v_{ij}$ if node $i$ is an end node of edge $e$, in which case node $j$ is the other end node. If node $i$ is not an end node of edge $e$, then $B_{ie} = 0$. The row corresponding to node $u$ is not included in $\mathbf{B}$. Because $\mathbf{M}$ can be written as the product of a matrix and its transpose, it immediately follows that $\mathbf{M}$ is positive semidefinite. To show that it is positive definite, we must prove that $\mathbf{B}^T \mathbf{x} = \mathbf{0}$ is true if and only if $\mathbf{x}$ is the zero vector. If $\mathbf{x}$ is the zero vector, then clearly $\mathbf{B}^T \mathbf{x} = \mathbf{0}$. Assuming $\mathbf{B}^T \mathbf{x} = \mathbf{0}$, let us consider the components separately. For an edge $e = (i,u)$ we have $(\mathbf{B}^T \mathbf{x})_e = x_i A_{ij} / v_{ij}$, which means that for all neighbours $i$ of node $u$, $x_i = 0$. For an edge $e = (i,j)$ ($i,j \neq u$) we have $(\mathbf{B}^T \mathbf{x})_e = (x_i + x_j) A_{ij} / v_{ij}$. If $i$ is a neighbour of $u$, then we already know that $x_i = 0$, therefore $x_j$ must also be zero. This condition propagates down to all nodes that are in the same connected component as $u$. We assumed that the underlying network is connected, therefore all components of $\mathbf{x}$ are zero, proving that $\mathbf{M}$ is positive definite.


\input{paper.bbl}

\end{document}

%% file: paper.bbl
%